\pdfoutput=1
\documentclass[journal]{IEEEtran}

\usepackage{graphicx}% Include figure files
\usepackage{dcolumn}% Align table columns on decimal point
\usepackage{bm}% bold math

\begin{document}
\title{Demonstration of a Dual Alkali Rb/Cs Atomic Fountain Clock}

\author{J.~Gu\'{e}na, %~\IEEEmembership{}
        P.~Rosenbusch, %~\IEEEmembership{}
        Ph.~Laurent, %~\IEEEmembership{}
        M.~Abgrall, %~\IEEEmembership{}
        D.~Rovera, %~\IEEEmembership{}
        M.~Lours, %~\IEEEmembership{}
        G.~Santarelli, %~\IEEEmembership{}
        M.E. Tobar, %~\IEEEmembership{}
        S.~Bize %~\IEEEmembership{}% <-this % stops a space
        and A.~Clairon %~\IEEEmembership{}
\thanks{Manuscript  received  June  3,  2009;  accepted  Cctober  30,  2009. 
This work  is  supported  by LNE. SYRTE  is  Unit\'e  Mixte  de recherche  du
CNRS (UMR 8630)  and  is  associated  with  Universit\'e  Pierre  et  Marie
Curie.}
\thanks{J.  Gu\'ena,  P. Rosenbusch,  Ph.  Laurent,  M. Abgrall, D.  Rovera,  G.
Santarelli, S.  Bize,  and  A.  Clairon  are  with  Laboratoire  National  de
M\'etrologie-Syst\`eme de  R\'ef\'erence Temps Espace (LNE-SYRTE), Observatoire de Paris, France (e-mail: jocelyne.guena@obspm.fr).}
\thanks{M. E. Tobar is with University of Western Australia,  School of Physics,  Crawley, Australia.}
\thanks{Digital  Object Identifier 10.1109/TUFFC.2010.1461}
}

\maketitle

\begin{abstract}
%\boldmath

We report the operation of a dual Rb/Cs atomic fountain clock. $^{133}$Cs and $^{87}$Rb atoms are cooled, launched, and
detected  simultaneously  in  LNE-SYRTE's  FO2  double  fountain. The dual clock operation occurs with no degradation of
either the stability or the accuracy. We describe the key features  for  achieving  such  a  simultaneous  operation.  We  also
report  on  the  results  of  the  first  Rb/Cs  frequency  measurement campaign performed with FO2 in this dual atom clock configuration, including a new determination of the absolute $^{87}$Rb hyperfine frequency.

\end{abstract}

%\begin{IEEEkeywords}
%Time and Frequency metrology, Primary frequency standards, atomic fountains, Fundamental constants, Cold atom space clock
%\end{IEEEkeywords}

% \fi
%\IEEEpeerreviewmaketitle

\section{Introduction}

The development of Rb atomic fountains was initially
motivated  by  the  prediction  that  the  cold  collision
shift in  Rb would be much reduced compared with Cs [1].
Following 2 independent experimental verifications of this
prediction [2], [3], the development of a high accuracy  Rb
fountain frequency standard has been pursued at SYRTE.
a first application of the $^{87}$Rb fountain was a measurement
of $\nu_{\mathrm{Rb}}$, the ground state hyperfine frequency of rubidium,
performed  at SYRTE  with  considerably  improved  accuracy [4]. Several $\nu_{\mathrm{Cs}}/\nu_{\mathrm{Rb}}$
 measurements followed, which all
involved  the  Rb  fountain  with Cs  reference  provided  by
the  SYRTE FO1 or the FOM mobile fountain. The $\nu_{\mathrm{Rb}}$ determination performed in 2002 was chosen as a secondary
representation of the  SI second by the Consultative Committee for Time and Frequency (CCTF) in 2004 [5]. The
main  interest  of  precise  and  repeated Rb/Cs  frequency
comparisons is to provide a test of variation of fundamental  constants  [6]–[8].  For  this  and  other  tests  involving
the alkali hyperfine frequency ratio, probing the 2 species
simultaneously  in  the  same  environment  was  considered
as a promising approach to improve the comparison. The
Rb/Cs dual fountain, dubbed FO2, developed at SYRTE
aims at pursuing this goal. Following longstanding developments of Rb and  Cs subsystems independently [7], [8],
the FO2 fountain operating with either Cs or  Rb as single
species at a given time proved among the most accurate
fountain clocks. Here we report on the first truly simultaneous operation of FO2 as a  Rb/Cs dual clock.

The key elements and techniques for dual mode operation are described in section II. section III deals with the
frequency accuracy of the dual clock. The first measurements performed with the dual clock FO2, and associated
results, are given in  section IV.

\section{dual Fountain  apparatus and Techniques}

\subsection{The Dual Atom Fountain Set-Up}

Fig. 1 shows the scheme of the Rb/Cs FO2 fountain.
The 2 atomic species are captured in the same region by
a  Rb/Cs  dual  optical  molasses  operating  in  a  lin $\perp$ lin
configuration. The Rb/Cs optical molasses are overlapped
using  dedicated  dichroic  collimators,  the  key  elements
for simultaneous operation. The laser beams at 780 and
852~nm, for cooling  Rb and  Cs, respectively, are generated
on 2 separate optical benches, then sent into the collimators via optical fibers and combined on a dichroic beam splitter (Fig. 2). The resulting beam is collimated using
a single achromatic lens to a diameter of about 26~mm.
The realization of these collimators [9] had to meet severe
constraints to ensure capture efficiency and verticality of
the  launch  direction  for  both Rb  and  Cs:  alignment  of
Rb  and  Cs  beams  to  $\sim 100~\mu$rad,  centering  of  intensity
profiles to $\sim 1$~mm, in addition to being non-magnetic and
removable to fit in existing magnetic shields. The 6 dual-wavelength laser beams are aligned along the axes of a 3-D
coordinate system, where the (1,1,1) direction is vertical.
The optical molasses are loaded from 2D-MOT presources
[10] for both  Rb and  Cs. Typical loading time is 500~ms
for both  Rb and  Cs, with total optical power of $\sim 100$~mW
using injection-locked laser diodes for Cs and $\sim 150$~mW
using  tapered  optical  amplifiers  for Rb.  The  2D-MOTs
decrease  the  consumption  of Rb  and  Cs  compared  with
previous  chirp  cooled  atomic  beams,  yet  at  the  price  of
decreased atomic fluxes. These presources also reduce the
background pressure in the capture zone.

Both species are launched upward at the same instant
with  a  slightly  different  velocity  of  $4.33$~m·s$^{-1}$ (apogee
0.96~m) for Cs  and  $4.16$~m·s$^{-1}$ (apogee  0.88~m)  for  Rb,
then cooled to $\sim$0.9 and 1.5~$\mu$K, respectively, in the  Cs  $F= $ and Rb  $F=2$ hyperfine ground states. There are 2
state-selection microwave cavities (at heights 64 and 139~mm above capture region for Cs and  Rb, respectively). For
Cs the  $F =3, m_F = 0$ initial clock state is populated with atom number adjusted for full or half density by adiabatic
population transfer [11]. For Rb, the  $F = 1, m_F = 0$ initial
clock state is populated by a microwave interaction from
the  $F = 2, m_F = 0$ state. Push beams (Fig. 1) throw out
Cs/ Rb atoms remaining in the unwanted atomic states ($F=4$ for Cs and  $F=2$ for  Rb).

The Ramsey interrogation probing the hyperfine transitions ($^{133}$Cs  $F = 3 \longrightarrow  F = 4$ at 9.192~GHz and
$^{87}$Rb  $F=1 \longrightarrow F=2$ at 6.834~GHz) consists of 2 microwave interactions ($\pi$/2 pulses) separated by a free atomic time of flight
for  each  species.  The  interactions  at  upward  and  downward  passages  take  place  in  a  special  double  microwave
cavity  made  of  2  resonators  on  top  of  each  other.  The height of the resonator center above the molasses zone is
0.442~m for Rb and 0.518~m for  Cs. The Rb/Cs resonators
are machined out of a single copper assembly (Fig. 3) for
compactness and to achieve temperature tuning for each
atomic species at the same temperature. Tuning for both
cavities  simultaneously  to  within  40~kHz  of  the  atomic
resonances  occurs  near  300~K.  The  resonators  are  $TE_{011}$
cylindrical  cavities  with  quality  factors  of  $\sim$~7100  for  Cs
and $\sim$~6000 for  Rb. Each resonator can be fed either symmetrically or asymmetrically using 2 opposite microwave
feedthroughs  allowing  the  study  and  reduction  of  first-order  doppler effects related to phase gradients inside the
cavity.

\begin{figure}[t]
\includegraphics [width=\linewidth]{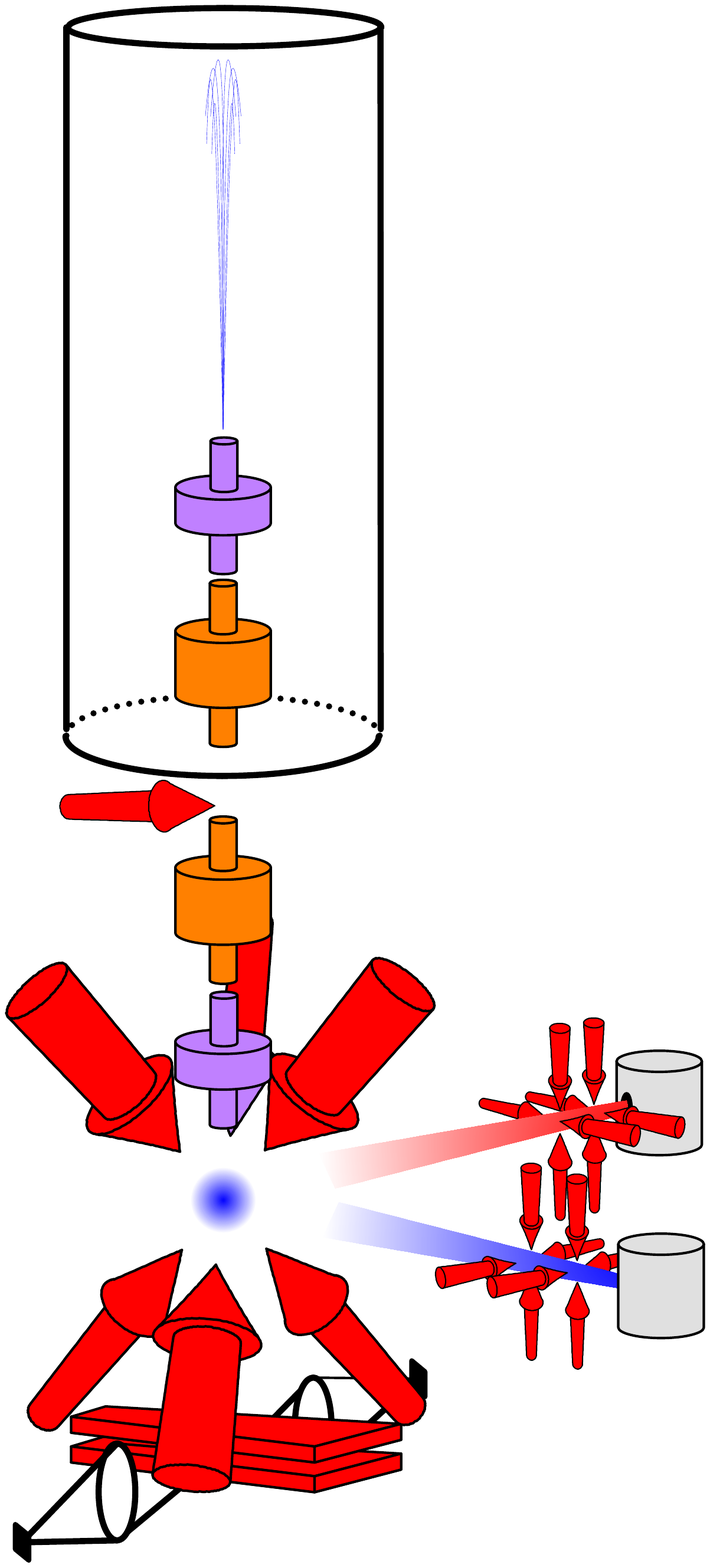}
\caption{Scheme of the dual fountain set-up.}
\end{figure}

\begin{figure}[t]
\includegraphics [width=\linewidth]{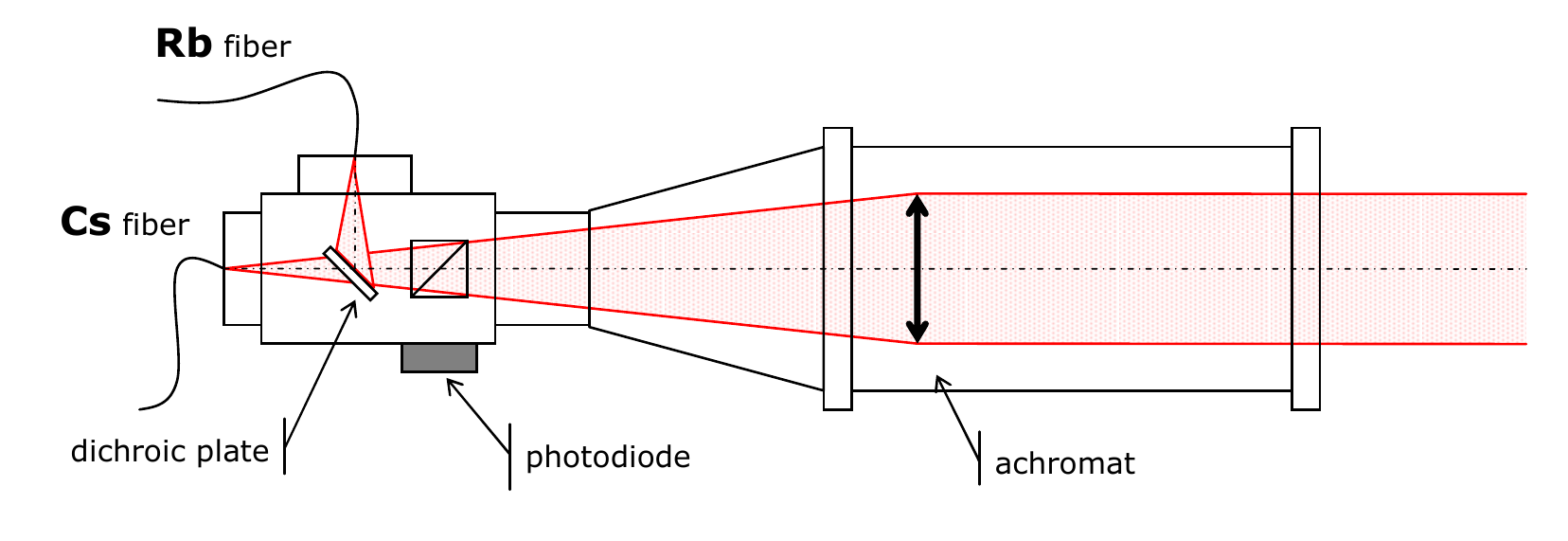}
\caption{Schematic of a dual collimator.}
\end{figure}
\begin{figure}[h]
\includegraphics [width=\linewidth]{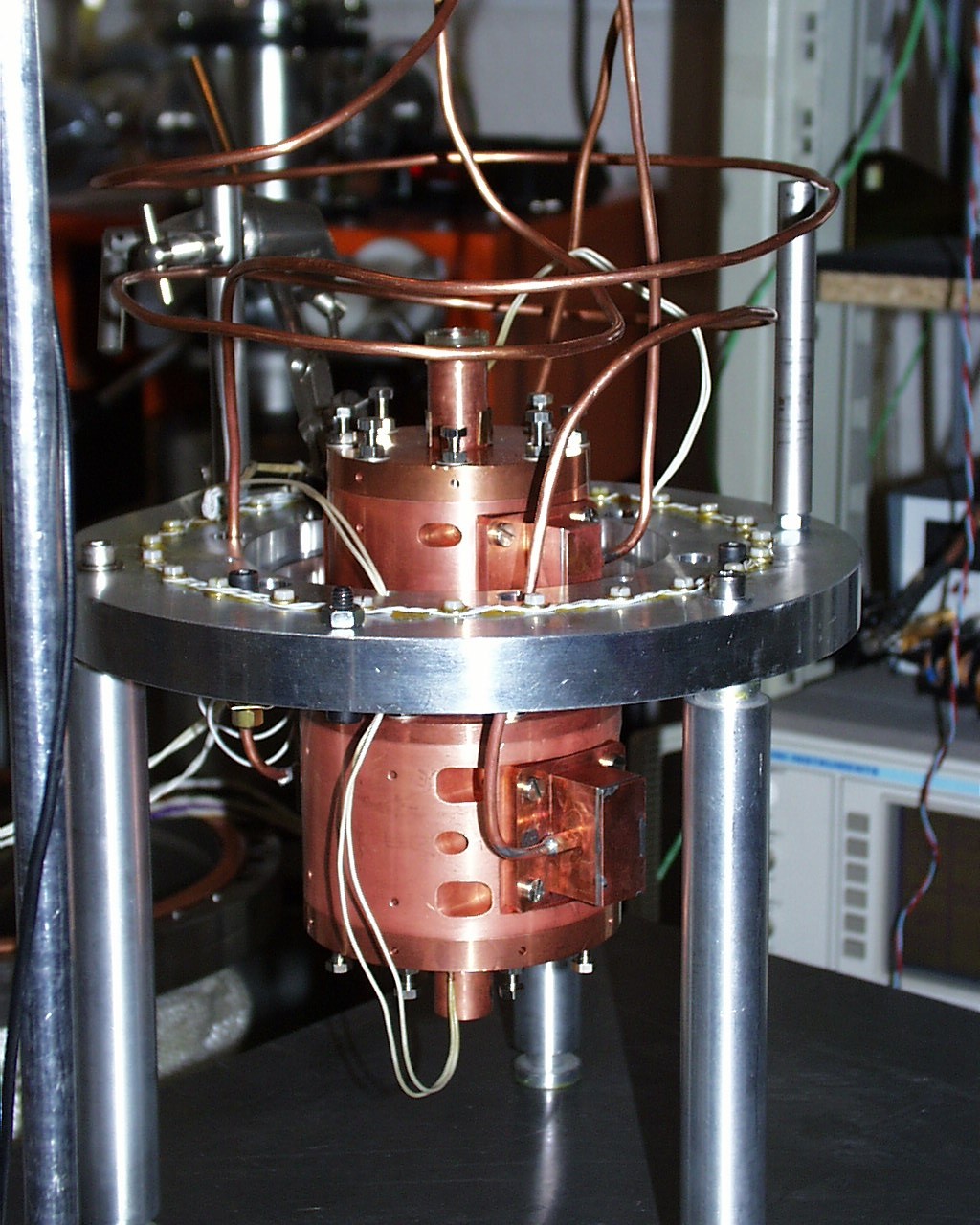}
\caption{The FO2 dual Rb/Cs  Ramsey cavity.}
\end{figure}

\subsection{Dual Detection Technique}\label{subsec_DDT}

The detection zone is located below the molasses zone.
It consists of 2 separate dual wavelength standing waves
of  resonant  light  (the  second  one  with  repumper  light)
allowing  detection  of  the  2  clock  states  for  each  atomic
species by induced fluorescence. The fluorescence light is
collected non-selectively onto the same photodetectors for
Rb and  Cs and selection between the 2 alkalis is temporal.
The  launch  velocities  are  chosen  to  avoid  collisions  between the 2 atomic clouds during the entire ballistic flight
(microwave state selection included), and then the launch
instants  are  finely  adjusted  so  that  there  is  no  overlap
between the  Rb and  Cs times of flight. The difference in
Rb/Cs  launch  instants  is  of  a  fraction  of  a  millisecond.
a plot of the dual  Rb/Cs ballistic flight is given in Fig.
4.  The  typical  dual  clock  cycle  is  1.5~s. note  that  the
Ramsey interrogation times, and hence the Ramsey fringe
widths,  are  equal  for  Rb  and  Cs  ($T=598$~ms,  Ramsey
fringe $\mathrm{FWHM}\cong 0.82$~Hz).

The fluorescence signal from one detection channel in
dual  mode  is  shown  in  Fig.  5.  despite  the  wings  of  the
cold  atom  velocity  distributions  (especially  for Rb),  the
overlap  between  Rb  and Cs  is  insignificant.  We  choose
to apply the detection beams for about 100  ms for the 2
atomic species consecutively without any temporal overlap.  In  these  conditions,  the  signal-to-noise  ratio,  and
hence the frequency stability, for each atomic clock alone
is preserved. The short-term frequency instabilities, limited by atom numbers, are at present about $3\times 10^{-14}$ at
1~s for Cs at high density and $5\times 10^{-14}$ at 1~s for Rb.

\begin{figure}[h]
\includegraphics [width=\linewidth]{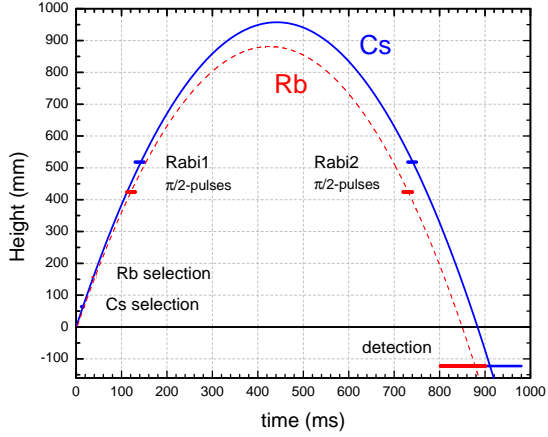}
\caption{Rb/Cs ballistic flights in dual-clock configuration. Time origin is
the launch instant and height origin is the molasses height.}
\end{figure}
\begin{figure}[h]
\includegraphics [width=\linewidth]{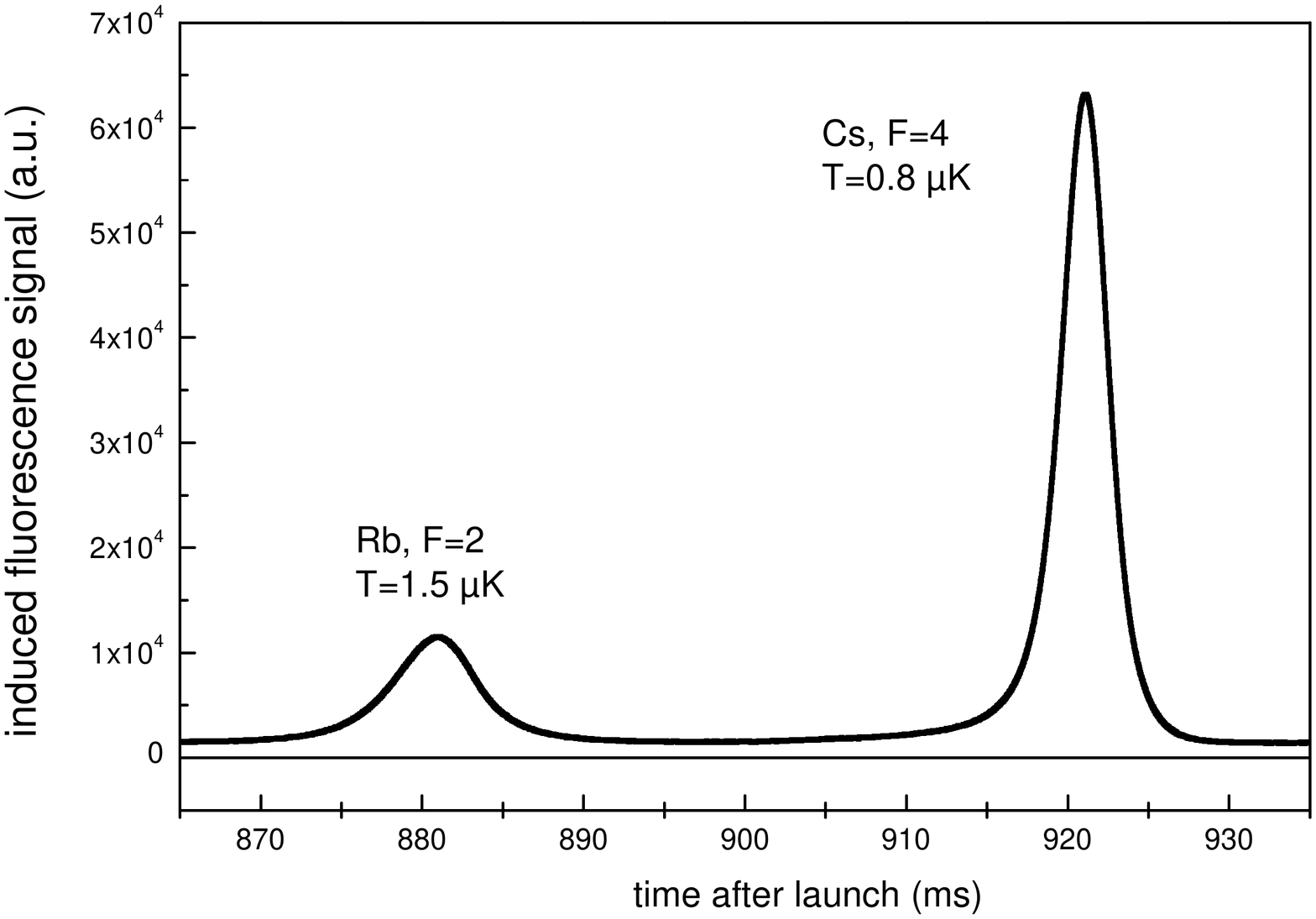}
\caption{Example of a dual fluorescence signal from the upper state detection channel with the time sequence given in Fig. 4. The interrogation
signals for both Rb and  Cs are at  Ramsey center fringe resonance of the
respective clock transitions, whereas for clock operation, the interrogation frequency alternates between the 2 sides of the fringe.}
\end{figure}

\subsection{Cs/Rb Synthesizers}\label{subsec_CSRbSynthe}

\begin{figure}[h]
\includegraphics [width=\linewidth]{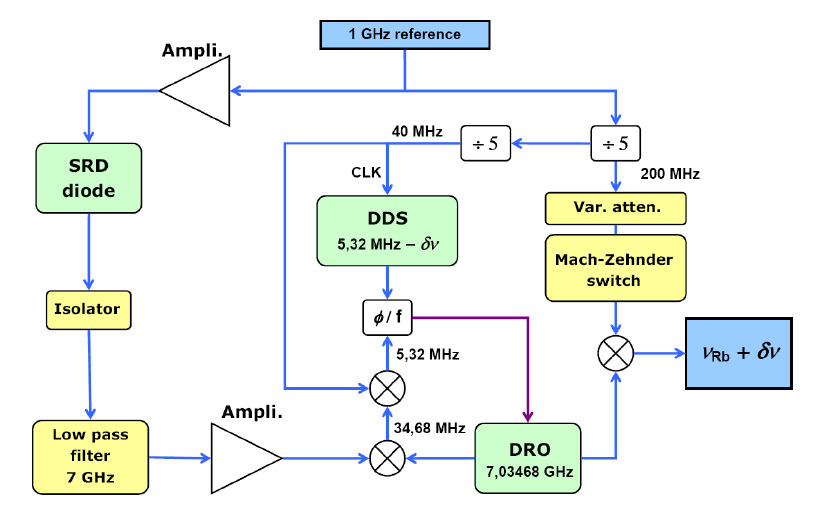}
\caption{Schematic of the FO2-Rb microwave synthesizer.}
\end{figure}
\begin{figure}[h]
\includegraphics [width=\linewidth]{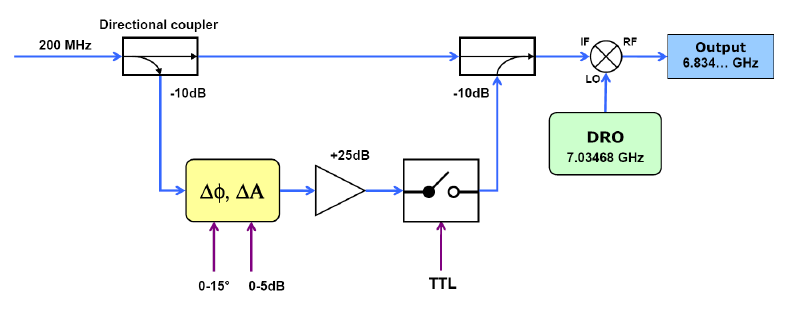}
\caption{The Mach-Zehnder switch in the FO2-Rb synthesizer.}
\end{figure}
The stability requirements of the frequency synthesizers driving the clock transitions are quite stringent, and
a variety of different designs have been implemented already. At LNE-SYRTE, the heart of the frequency synthesis is a cryogenic sapphire oscillator (CSO) operating near 11.932~GHz. a frequency offset stage already described in
[12] shifts its output signal to 11.98~GHz with tunability
using a direct digital synthesizer (DDS). The 11.98~GHz
signal reproduces the low phase noise and high short-term
frequency stability of the  CSO. This highly stable signal
is down converted to 1~GHz and to 100~MHz for distribution  purposes  [13].  To  compensate  the  drift  of  the CSO
($\sim 10^{-18}/\mathrm{s}$ in relative frequency), the signal at 11.98~GHz
is phase-locked to the 100~MHz output signal of a H-maser
(with time constant of 1000~s).

In the present configuration, the Cs synthesizer uses the
low phase noise 11.98~GHz signal to generate 9.192~GHz
in a home-built frequency chain, whereas the 1~GHz signal
is up-converted to 6.384~GHz for Rb as shown in Fig. 6.
Each frequency synthesizer incorporates a computer-controlled DDS digital synthesizer with microhertz resolution
to  tune  the  interrogation  signal  to  the  clock  transition.
The  frequency  corrections  applied  to  the  DDSs  are  the
basis for evaluating the frequency stability and frequency
shifts.  Each  synthesizer  includes  a  Mach-Zehnder  interferometer  RF  switch  at  400~MHz  or  200~MHz  (Fig.  7).
The  switch  provides  an  extinction  ratio  of  60  to  70~dB
of  the  microwave  signal  without  introducing  significant
phase transients [14], [15].  during the clock operation, the
switch turns off the microwave signal when atoms are outside the microwave cavities with the purpose of suppressing possible microwave leakage during the  Ramsey inter-rogation. The phase noise power spectral density and the
long-term stability of the synthesis have been extensively
tested. The noise is negligible at the $10^{-14}\tau^{-1/2}$ level. Each
synthesizer has 2 outputs with power and phase adjustments on one channel to provide symmetrical feeding of
the  Ramsey resonators.

\section{Systematics/Accuracy Budgets}

\begin{table*}
\renewcommand{\arraystretch}{1.3}

\begin{center}
\begin{tabular}{lccc}
\hline
 & {\small  FO2-Cs}& {\small  FO2-Rb} &{\small  FO1-Cs}\\
 & $\mathrm{corr.}\pm \mathrm{unc.} (\times 10^{-16})$ & $\mathrm{corr.}\pm \mathrm{unc.} (\times 10^{-16})$ &$\mathrm{corr.}\pm \mathrm{unc.} (\times 10^{-16})$\\
\hline
{\small Quadratic Zeeman effect}                & $-1914.3 \pm 0.3$         & $-3468\pm 0.7$            & $-1276.7\pm 0.2$     \\
%\hline
{\small Blackbody radiation}                    & $167.2 \pm 0.6$           & $120.6\pm 1.6$            & $165.2\pm0.6$      \\
%\hline
{\small Cold collisions and cavity pulling}     & $246\pm 2.5$              & $8.4\pm 2.6$              & $81 \pm 2.2$     \\
%\hline
{\small First-order Doppler effect}             & $0\pm 3$                  &$0\pm 2.5 $                &$0\pm 3.2$                  \\
%\hline
{\small Microwave leakage \& spectral purity}   & $0\pm 0.5$                & $0\pm 0.5$                &$0\pm 1$                   \\
%\hline
{\small Others (quantum motion,}&&\\
{\small background gas collisions,}&&\\
{\small Ramsey \& Rabi pulling, etc.)}          &$0\pm 2.0$                 &$0\pm 2.0$                 &$0\pm 1.7$               \\
%\hline
{\small Total uncertainty}                      & $4.4$                     & $4.5$                     & $4.4$            \\
 \hline
\end{tabular}
\end{center}
\caption{Accuracy budget for SYRTE FO2-Cs/Rb and FO1 fountains.}
\label{tab_accuracy}
\end{table*}

Table I gives the budget of systematic frequency shifts
and associated uncertainties for FO2-Cs/FO2-Rb in dual
clock operation. By far the largest systematic effect is the
quadratic Zeeman shift and Rb is twice as sensitive as  Cs.
Great care has to be taken to measure the magnetic field
(C-field) value. For this purpose we use  Ramsey spectroscopy  on  the  first-order  Zeeman  sensitive  transitions  for
both Cs and  Rb. Fig. 8 shows the  C-field maps obtained
with Rb and  Cs consecutively. Note that each point in this
figure is the average magnetic field over the atom trajectory at different launch heights. Because the Rb/Cs microwave cavities are not at the same height (height difference is 76~mm) the 2 maps are not perfectly identical. The C-field of about 2~mG is homogeneous at the low $10^{-3}$ level,
leading to a relative frequency uncertainty less than $10^{-18}$
during clock operation at standard apogees (of 0.88~m for
Rb and 0.96~m for Cs) we periodically check the stability
of the  C-field. The quoted uncertainty in Table I accounts
for temporal fluctuations and statistical uncertainty in the
field measurement, as well as uncertainties in the Zeeman
coefficients involved.

The uncertainty in the blackbody radiation frequency
shift (BBR) is larger for Rb compared with  Cs because 1)
the theoretical estimate has not been carried out with as
much precision as for  Cs [16], and 2) no measurement of
the  Rb  stark coefficient has been performed with similar
accuracy  as  for Cs  [17].  The  cold  collisions  shift  is  extrapolated in real time for  Cs exploiting interleaved measurements at high/low Cs density using adiabatic passage
[11].  The  technique  is  not  implemented  for  Rb  because
the  related  effect  is  much  smaller  (by  a  factor  of  about
30) with the present atom number. Instead, we perform
clock measurements alternating between 2 loading times
(600  and  300~ms)  for Rb. note  that  the  atom  number
dependent shift includes both contributions from cold collisions and from cavity pulling due to a small cavity detuning ($\sim 250$~kHz). The associated uncertainty is mainly statistical. In the future we plan to systematically perform
such interleaved measurements for Rb, just as for  Cs. This
will  be  all  the  more  desirable  if  we  succeed  in  increasing the  Rb atom number (by increasing optical power) to
improve  the  Rb  clock  stability.  at  present,  the  limiting
systematic uncertainty is the first-order  Doppler effect associated with cavity phase variations. Adjustment of the
symmetric feeding of the cavities and of the launch direction provides effective reduction of the odd terms in the
phase distribution. The quoted uncertainty in Table I is
based on theoretical estimates of even terms of distributed
cavity  phase  (DCP)  [18],  [19].  Work  is  ongoing  on  both
experimental and theoretical sides to reduce the DCP un-certainty toward the $10^{-16}$ level.

An extended  search  for  microwave  leaks  in  the  FO2-Rb/Cs fountain has been conducted using the microwave switch [14], [15]. Differential measurements with continuous or pulsed application were performed at different microwave powers. no frequency difference at the $10^{-16}$ level
at the nominal power of $\pi/2$ pulses could be concluded.
The leakage effect is much smaller during regular operation with microwave switched off (extinction ratio $>$50 to
60~dB) during the free time of flight.

In Table I, the overall accuracy is the quadratic sum of
all systematic uncertainties. The budget for the  SYRTE FO1-Cs fountain is also given. This fountain was involved
in the measurements presented in next section.
\begin{figure}[h]
\includegraphics [width=\linewidth]{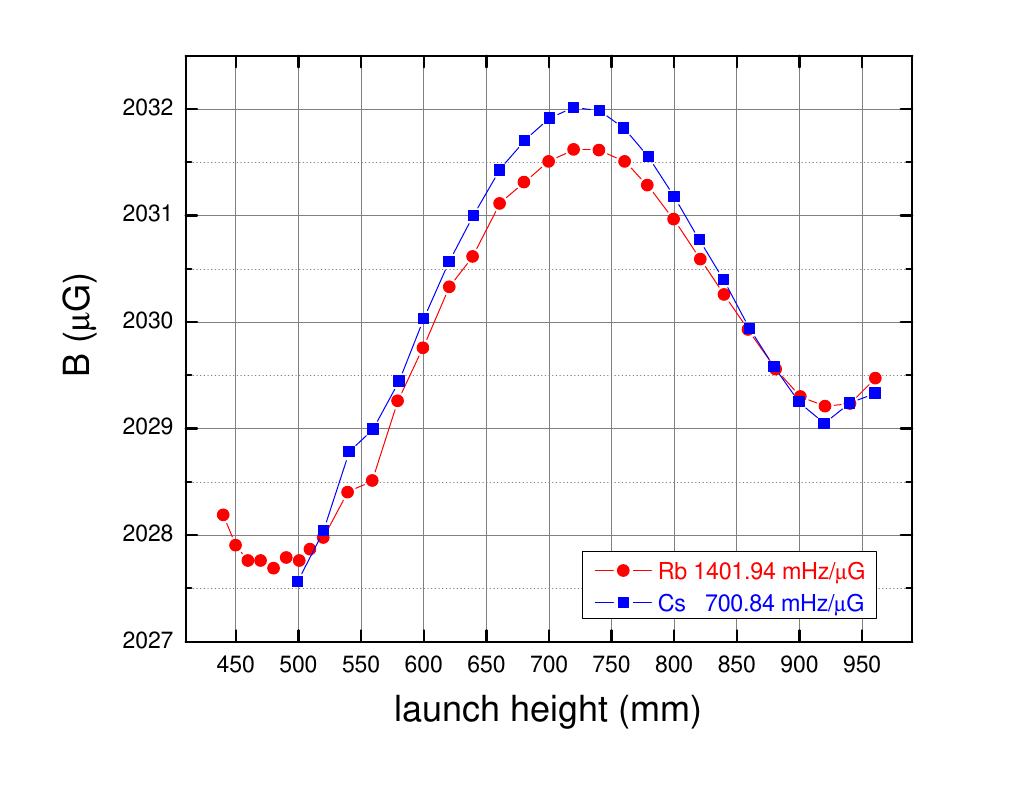}
\caption{C-field maps in the interrogation region probed by the FO2-Rb
and FO2-Cs atoms on the Zeeman 1–1 transitions.}
\end{figure}

\section{First  Rb/Cs Frequency Measurements With
the dual atom clock FO2}

\subsection{Rb-Cs Differential Frequency Fluctuations}

\begin{figure}[h]
\includegraphics [width=\linewidth]{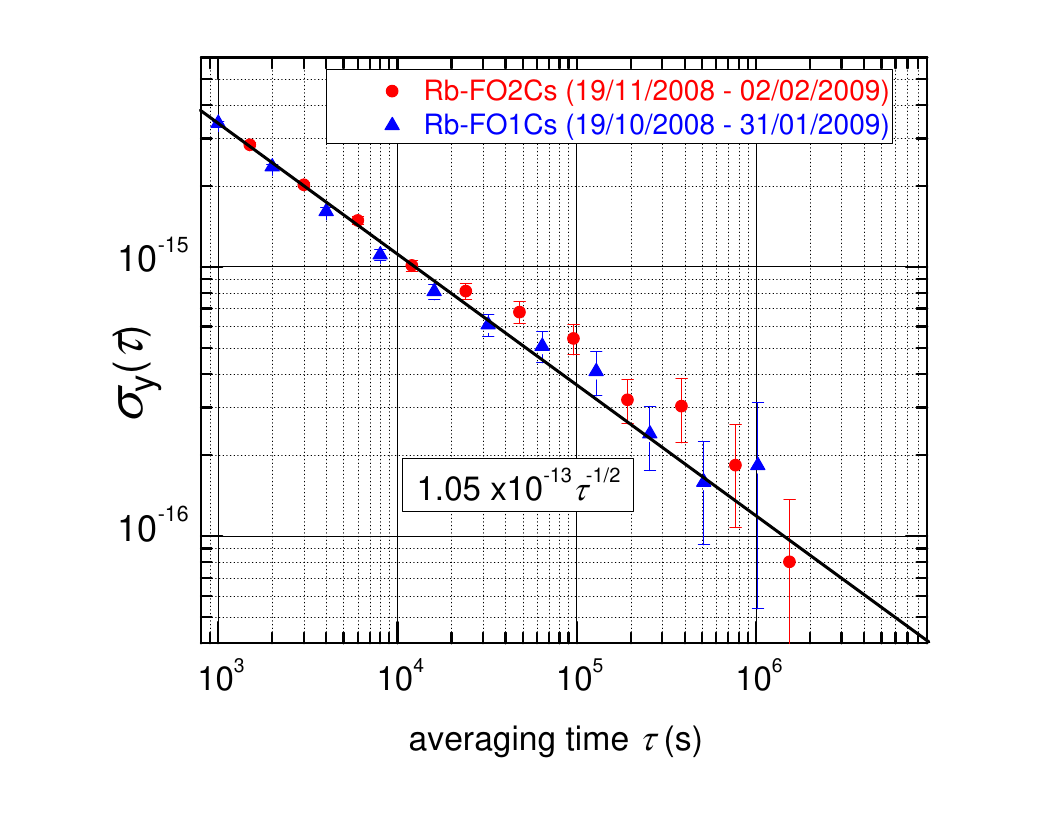}
\caption{Allan standard deviation for the Rb/Cs relative frequency differences measured with FO2 as dual alkali clock. Circles: FO2-Rb/FO2-Cs
comparison, triangles: FO2-Rb/FO1-Cs comparison.}
\end{figure}
The first measurement campaign with FO2 operating
as a dual alkali atom clock took place from November 19,
2008  to  January  31,  2009,  covering  48  effective  days  of
data. Fig. 9 presents the allan standard deviation for the
relative  frequency  difference  between  FO2-Cs  and  FO2-Rb  (full  squares). Data  are  corrected  for  all  systematic
effects except for the shift dependent on  Rb atom number
which  is  corrected  in  post  data  processing  (see  section
III). assuming  white  noise,  the  statistical  resolution  of
about $10^{-16}$ is reached in 20 days.

During the same period, FO1-Cs fountain was also running.  The  full  circles  in  Fig.  4  represent  the  allan  deviation  for  the  FO2- Rb/FO1  frequency  comparison  over
synchronous data (36 effective days of data). It turns out
that  both  comparisons  have  rather  similar  noise  behavior. Both indicate a good rejection of the local oscillator
noise: indeed, the relative frequency drift of the H-maser
(of typically a few times $10^{-16}$
 per day) which is conspicuous for a single clock is clearly absent in the comparison
for either pair of clocks.

\subsection{New $^{87}$Rb Hyperfine Frequency Determination}

\begin{figure}[h]
\includegraphics [width=\linewidth]{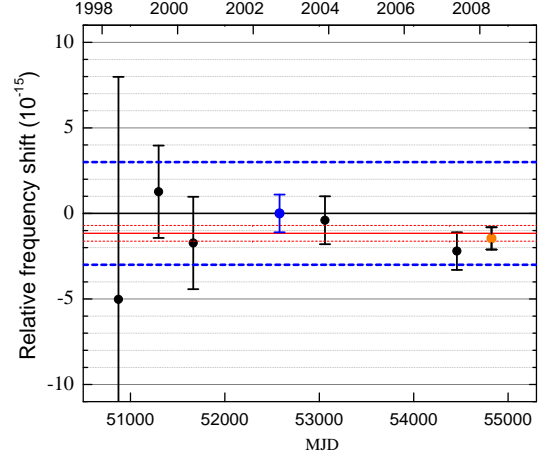}
\caption{Relative hyperfine frequency of $^{87}$Rb relative to that of
$^{133}$Cs as a function of time. The data point at MJD 52579 chosen as the frequency
reference  is  the  value  recommended  by  CCTF  2004.  The  thick  dotted
blue lines correspond to the recommended uncertainty. The red continuous line having an offset is the result of a weighted least square fit to all
the data by a constant value. The thin dotted red lines correspond to the
fit uncertainty at 1 $\sigma$.}
\end{figure}
The  present  Rb-Cs  double  campaign  provides  a  new
value for the $^{87}$Rb hyperfine frequency. We take as reference  the  value  recommended  by  the  CCTF  in  2004  [5], which  is  the  value  measured  at  LNE-SYRTE  in  2002.
Then,  from  the  FO2-Rb/FO2-Cs  comparison,  we  obtain
the mean Rb-Cs hyperfine relative frequency difference:
\begin{equation}
    \begin{array}{cc}
    \nu(\textrm{FO2-Rb})(2009)/\nu_{\mathrm{Rb}} (\textrm{CCTF}) -1\\
    =(-1.68 \pm 0.65) \times 10^{-15},
    \end{array}
\end{equation}
and from the FO2-Rb/FO1-Cs comparison,
\begin{equation}
    \begin{array}{cc}
    \nu(\textrm{FO2-Rb})(2009)/\nu_{\mathrm{Rb}}(\textrm{CCTF}) -1\\
    =(-1.23 \pm 0.65) \times 10^{-15}.
    \end{array}
\end{equation}
The 2 results are consistent within the error bars, which
are dominated by the systematic uncertainties (Table I).
We choose to average them with equal weights. This yields
the absolute $^{87}$Rb hyperfine frequency
$$\nu(\textrm{FO2-Rb})(2009) = 6~834~682~610.904314(4)~\mathrm{Hz}.$$
The recommended value
$$\nu_{\mathrm{Rb}}(\textrm{CCTF-2004}) = 6~834~682~610.904324(21)~\mathrm{Hz}$$
within  the  assigned  relative  uncertainty  of  $3\times 10^{-15}$  is
consistent with the new determination.

\subsection{Overview of Rb-Cs Frequency Comparisons}

Fig. 10 shows the results of the Rb-Cs frequency measurements performed at  LNE-SYRTE since 1998. For the
first 6 points, FO2-Rb was operating as a single clock and
the Cs reference was FO1, the FOM mobile fountain, or
both. The present result with FO2 as dual  Rb/Cs clock
(last  point)  is  perfectly  consistent  with  previous  results. A weighted least squares fit to the data by a constant has
a  $\chi^2$-goodness  of  fit  probability  of  $Q=0.75$,  indicating
internal consistency of the data (weighted mean $-1.16\times 10^{-15}$, uncertainty $0.46\times 10^{-15}$ at 1 $\sigma$).

\section{Summary and Outlooks}

In this paper, we have presented the first demonstration
of a dual Rb/Cs alkali clock with the 2 species simultaneously probed in the same device. As it is implemented, the
simultaneous  clock  operation  preserves  the  stability  and
accuracy  of  each  clock  operated  alone.  It  also  preserves
the  possibility  of  configuring  the  2  clocks  independently
for a large variety of measurements.

To demonstrate  the  capability  of  this  setup,  we  have
used  the  dual  fountain  to  perform  a  new  absolute  frequency measurement of the Rb hyperfine frequency. This
measurement has a slightly improved uncertainty and is
consistent  with  all  measurements  performed  before  the
implementation  of  the  simultaneous  operation  with  Rb
and  Cs.

For long-term operation, for  Rb/Cs comparisons over
extended period of time and their application to testing
the stability of fundamental constants, the dual fountain
setup  provides  a  practical  gain  of  being  a  single  device
in  a  single  room  with  a  single  operator.  Moreover,  the
collocated  interrogation  of  both  species  allows  a  partial
cancelation of some systematic shifts in the measurement
of the  Rb/Cs frequency ratio. For instance, in fractional
terms, the  Rb/Cs ratio is $\sim 2.6$ times less sensitive to the
blackbody radiation shift than  Rb or Cs alone. It is $\sim 2$
times less sensitive to the second-order Zeeman shift than
Rb alone.  Also, temporally matched interrogation of both
atoms allows common-mode rejection of the interrogation
oscillator phase noise. This can lead to large improvement
when  the  stability  is  limited  by  the  oscillator  noise,  as
shown  in  [20].  However,  this  advantage  no  longer  exists
when  quantum  noise  limited  stability  is  achieved  by  using  an  ultra  low  noise  interrogation  signal  as  delivered
by  a  cryogenic  sapphire  oscillator  [21]  or  a  femtosecond laser [22].

The dual-fountain configuration also opens the way to
several new measurements. For instance, a modified version of the  lorentz invariance test performed with FO2-Cs
[23] could be performed using  Rb and  Cs simultaneously.
The use of  Rb together with Cs would extend the test to a
different sector of the theoretical framework. Furthermore,
magnetic field fluctuations which were limiting the stability of the measurements reported in [23] would be strongly
eliminated by common-mode rejection. The dual fountain
configuration  also  opens  the  possibility  of  studying Rb/Cs cold collisions. For time and frequency metrology applications, the time sequence is carefully chosen to avoid Rb/Cs collisions. However, the FO2 fountain set-up offers
flexibility  in  the  Rb/Cs  time  sequences.  Parameters  can
be easily adjusted to finely tune collision energy between
the 2 clouds.

%\cite{Kokkelmans1997,Fertig2000,Sortais2000,Bize1999,CCTF2004,Marion2003,Bize2004,Bize2005,
%Chapelet2008,Dieckmann1998,Pereira2002,Chambon2005,Chambon2007,santarelli2009,Guena2008,Angstmann2006,Simon1998,Li2005,Schroder2002,Bize2000a,
%Santarelli1999,Millo2009a,Wolf2006}

%\section*{Acknowledgments}

%\bibliographystyle{IEEEtran}
%\bibliography{D:/Bibliographie/BiblioLatex/bibliography}

\end{document}